# Optically Rewritable Memory in a Graphene–Ferroelectric-Photovoltaic Heterostructure

D. Kundys,[1] A. Cascales,[2] A. S. Makhort,[2] H. Majjad,[2] F. Chevrier,[2] B. Doudin,[2] A. Fedrizzi,[1] and B. Kundys[2,*]

[1]*SUPA, Heriot-Watt University, Edinburgh, EH14 4AS, UK*
[2]*Université de Strasbourg, CNRS, Institut de Physique et Chimie des Matériaux de Strasbourg, UMR 7504, 23 rue du Loess, 67000 Strasbourg, France*

Achieving optical operation of logic elements, especially those that involve two-dimensional (2D) layers, can begin the long-awaited era of optical computing. However, efficient optical modulation of the electronic properties of 2D materials, including the rewritable memory effect, is currently lacking. Here we report a fully optical control of the conductivity of graphene with write-erase operation yet under ultralow optical fluence. The competition between light-induced charge generation in a ferroelectric-photovoltaic substrate and relaxation processes provides the selective photocarrier-trapping control affecting the doping of the 2D overlayer. These findings open the way to photonic control of 2D devices for all-optical modulators and a variety of all-optical logic circuits, memories, and field-effect transistors.

## I. INTRODUCTION

One of the key challenges of today's technology is to overcome the slowing down of Moore's law for circuit miniaturization, which is approaching the single-atom level [1]. The development of new device paradigms is therefore crucial. In that respect, single-atom layer structures with unique electronic properties [2] allow atomic scale control [3] that combines electronic and photonic degrees of freedom. In particular, the high sensitivity of two-dimensional (2D) layers to the presence of nearby electric charges can be exploited in their combination with electric-charge-tunable environments. The most-successful materials in that respect are ferroelectric (FE) compounds, which provide electric-field-driven large-scale changes in charge doping related to bistable 180° polarization switching [4–14] or a variety of possible intermediate states that are associated with the existence of a FE-domain structure [15–17]. As the demand for operating speeds approaches photonic regimes [18,19], the electric control, however, must be replaced by optical control to enable fast and low-power-consumption data processing [19]. To accomplish this transition fully, current integrated photonics urgently needs to achieve optical functionalities that can efficiently interface photonic and electronic degrees of freedom for active and passive elements performing data processing and photon routing for memory read-write integrated elements [20–22].

The photopolarization properties of FE compounds also make them exceptional components for 2D hybrid structures, where large changes in charge density can be induced optically. Indeed, the possibility of optical control in 2D–FE hybrid structures has recently gained increasing interest [23–27]. However, to realize the full potential of the photopolarization properties of 2D–FE heterostructures, one should also take into account the possible existence of the photovoltaic effect in ferroelectrics [28–34]. The photopolarization in this case can additionally benefit from free photocarriers, as long as the light-excitation energy is close to the band gap. It is important to note that, in contrast to the slower thermal-pyroelectric effects, photopolarization of photovoltaic origin offers ultrafast (terahertz) operation [35,36] and can consist of both reversible and irreversible parts [37,38]. Such effects can offer unprecedented functionalities for all-optical control not only preserving nonvolatility, which is indispensable for memories, but also offering optical tunability of electronic and plasmonic properties in 2D–FE-heterostructure devices. In this work repetitive optical writing and optical erasing of graphene resistive states in a graphene overlayer based on the photopolarization response of the photovoltaic-ferroelectric substrate. The distinct physical mechanism involves delicate interplay between screening of remanent polarization produced by the photogenerated carriers, on the one hand, and the amount of trapped charges, on the other hand. Such a demonstration of both writing and erasing operation of graphene resistive states performed all optically is much needed, but seems to be

*kundysATipcms.fr

absent from the literature. Our approach also allows an all-optical 87% conductivity change in a 2D graphene layer obtained totally by optical means with exceptional efficiency of the graphene light-matter interaction (fluence 6 mJ/cm$^2$). Furthermore, the observed rate of conductivity change is at least 3 orders of magnitude faster than previously reported [25] and has potential for further improvement. This method opens the prospect for ultraefficient, all-optically-controlled 2D–FE devices incorporating electronic and photonic functionalities for memories, integrated photonics, sensors, and other optoelectronic applications, including a whole variety of possible combinations. More generally, giant all-optical switching of graphene's conductivity is desired for efficient memory devices and integrated photonics, where graphene plays the role of an ultrathin waveguide and other optoelectronic applications for light emission, transmission, modulation, and detection.

## II. EXPERIMENTAL DETAILS AND ELECTRICAL GROUND STATE

Figure 1 (left inset) illustrates the experimental approach used for electric-field-induced remanent-state formation. The Pb[(Mg$_{1/3}$Nb$_{2/3}$)$_{0.70}$Ti$_{0.30}$]O$_3$ crystal with dimensions of 2.43 × 1.09 × 0.26 mm$^3$ is cut from plates of (001) orientation supplied by Crystal GmbH (Germany). The graphene monolayer is deposited by CVD transfer on the top of the (001) plane without postannealing. The drain and source Au electrodes of 30-nm thickness are deposited on the graphene layer by electron-beam evaporation and wired with silver conductive paste. The resistance is measured with Agilent *LCR* meter at 100 kHz under ambient conditions. The FE loop is measured in darkness with a quasistatic loop tracer similar to that described in Ref. [39] by ultralow-frequency (0.01-Hz) measurements. We use above-band-gap excitation [40] provided by a 365-nm light-emitting diode (3.4 eV) with 30-nm spectral linewidth to generate free carriers in the Pb[(Mg$_{1/3}$Nb$_{2/3}$)$_{0.70}$Ti$_{0.30}$]O$_3$ substrate [Fig. 2(a)]. The Au electrodes are covered during illumination to prevent possible charge injection [41].

The ground state of the structure is first characterized electrically to verify ferroelectricity and ferroelectrically-driven performance before optical excitation. The FE polarization of the substrate (Fig. 1, right inset) leads to charge-driven modification of the graphene electrode forming also the two remanent states of resistance (Fig. 1). The typical antihysteretic behavior with an electric field is observed, indicating the possible presence of water at the interface between graphene and the FE [9,11,14,42]. This adds a background doping to the graphene overlayer, shifting the Dirac point. Therefore, the maximum resistance occurs at higher electric fields than the ferroelectric coercive force of the substrate (Fig. 1, right inset), where polarization should pass through zero. The optical excitations are performed at zero electric field in the well-defined remanent state (state 1) obtained after a positive poling.

## III. OPTICAL MEMORY EFFECT

Because of the possible existence of multiple polarization states in ferroelectrics [43] the electric ground state should be precisely defined and stabilized before the optical excitation. To this end, the structure is electrically poled in darkness to the remanent FE state by sweeping electric field from 0 to −7 kV/cm followed by a sweep from −7 to 7 kV/cm and then back to 0 kV/cm (point 1 in Fig. 1). Time-dependent measurements of the electric polarization are then conducted to ensure polarization stability before the optical excitation (Fig. 2). Sample illumination along the [001] direction [Fig. 2(a)] results in the large change of graphene resistance reaching 87% even with an extremely weak light power of 75 $\mu$W (fluence 6 mJ/cm$^2$) [Fig. 2(b)], greatly exceeding previous observations [23–26]. The light illumination decreases the electric field in the sample and hence the polarization [Fig. 2(b), lower part] due to the screening by the photogenerated charges. The dynamics of the photoresponse of the graphene consists of a faster contribution leading to a decrease in the resistance and a slower contribution leading to resistance increase [Fig. 2(b), inset]. It is worth noting that the response, which occurs on a subsecond timescale (approximately 200 ms), relates to the limit of our time resolution for resistive measurements and may have room for significant improvement. In particular, by changing the FE substrate to, for example, BiFeO$_3$ [35,36] or optimizing its thickness, a significant reduction in response time can be expected.

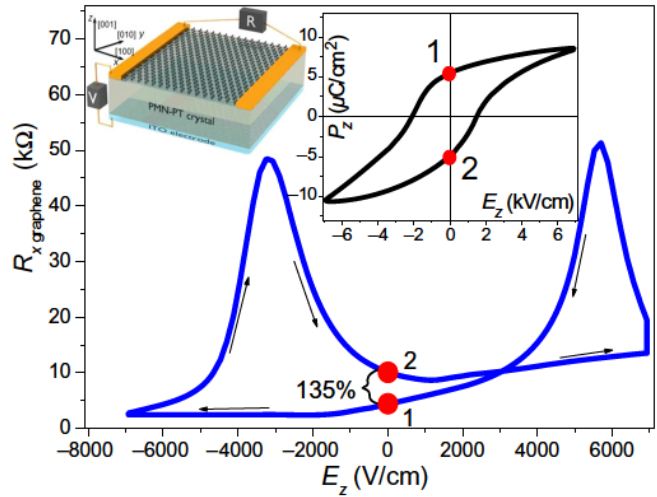

FIG. 1. An electric field induces evolution of the graphene resistance due to electrostatic doping. The top-right inset shows the related FE loop of the substrate. The top-left inset shows the experimental setup where a 2D graphene layer acts as an electrode of the FE capacitor.

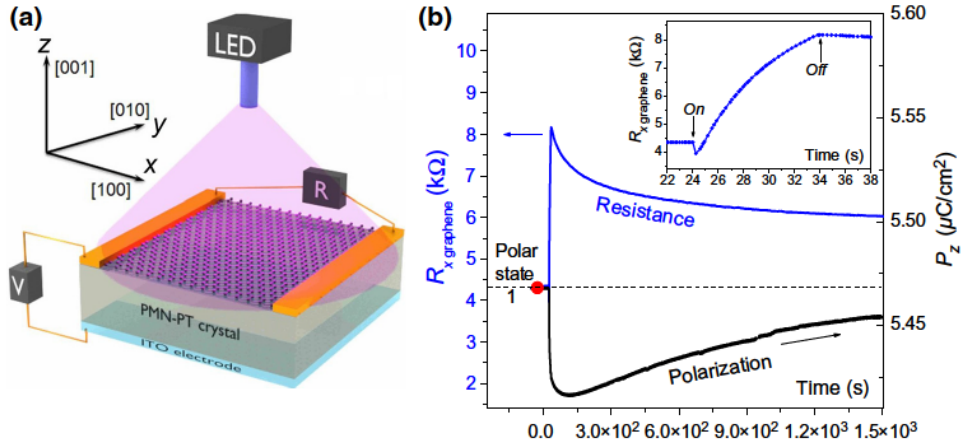

FIG. 2. (a) Experimental setup for optical excitations; the 2D graphene layer is illuminated by a 365-nm light-emitting diode along the [001] direction of the FE substrate, previously subjected to positive poling (state 1). (b) Polarization along the $z$ axis and resistance of the graphene layer in response to a 10-s light pulse of extremely weak intensity (0.60 mW/cm$^2$). The inset shows an enlargement of the photoresistive response indicating the nonlinear behavior.

When the light is switched off, the change in polarization of the substrate experiences relaxation [Fig. 2(b), lower part] but does not return to the initial level, constituting an optically written state that can be erased electrically by poling the system back to point 1 again, thus enabling a functionality of repeated optical writing and electrical reset. The increase of graphene resistance under illumination correlates with a decrease of remanent polarization, in agreement with Fig. 1(b), and is attributed to the photo(de)polarization effect [37]. The small decrease in the remanent polarization (approximately 1% of $P_z$ in this case) and uniform sample illumination rule out the formation of FE domains due to light in our initially monodomain sample. In the enlargement of the graphene photoresponse [Fig. 2(b), inset] the initial resistance decrease and the subsequent increase suggests a strong nonlinear mechanism involved in the observed phenomenon. To gain insight into the origin of the observed effect, the resistance is measured as a function of light intensity (Fig. 3).

The photogenerated electron-hole pairs distribute themselves along the previously formed polarization direction [001], allowing the graphene layer to be doped. This doping can be reversed electrically by our moving the system to point 2 [Fig. 1(b)], which agrees with our model of the photodepolarization effect described in detail in Ref. [37]. This light-induced charge-generation process occurs at intensities below 20 mW/cm$^2$ (Fig. 3) and eventually reaches saturation at higher intensities when the excess photocarriers start to recombine, thus reducing the remanent polarization, which is reflected in the resistance readout of the graphene overlayer. Because some of the charges have recombined irreversibly [37,38,44] a repeatable optical writing of the graphene resistive state is possible.

Moreover, it is clear that one can control the state formation by simply controlling the amount of trapped charges.

## IV. OPTICAL WRITING AND ERASING OF GRAPHENE RESISTIVE STATES

As seen from Fig. 3, light-excitation intensity below 20 mW/cm$^2$ leads to a decrease in graphene resistance, while intensities above this value have the opposite effect. Consequently, a balance can be found between the two processes by controlling the amount of trapped charges and their re-excitation. This can be done by varying the light fluence and illumination history [Fig. 4(a)].

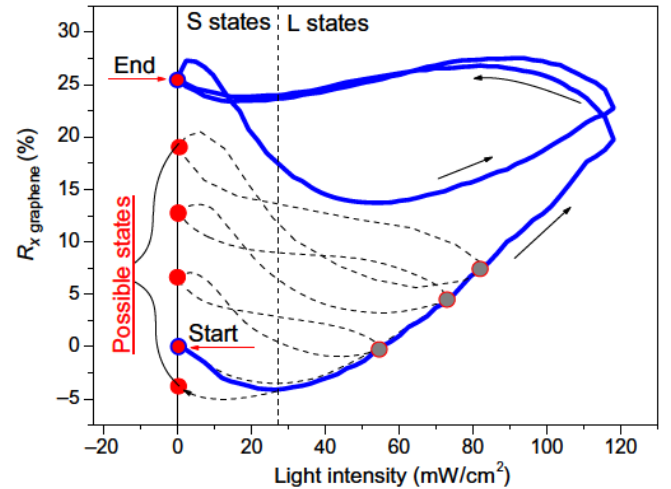

FIG. 3. Light-intensity sweeps from 0 to 118 mW/cm$^2$ (back and forth), with a step of 2.36 mW/cm$^2$, with the time step of 200 ms. Multiple light-written states of S and L types are possible depending on the driving-light intensity owing to the selective charge-trapping approach (see the text).

Depending on the illumination profile, the graphene states with larger (L) and smaller (S) photoresistive responses can be formed [Fig. 4(a)]. For comparison purposes, all states indexed from 1 to 4 are formed with the incremental steps of the same intensity as in Fig. 3. The other states are formed by light pulses with specific profiles (Fig. 4, lower plots), demonstrating the possibility to increase the writing speed. While the states of the L type show a tendency for relaxation, states of the S type exhibit significant stability. It is also clear that one can operate between states of the S type is possible by use of a specific illumination history, in agreement with Fig. 3. This possibility is illustrated in detail in Fig. 4(b), where squarelike writing pulses of higher intensity are followed by stabilization pulses of lower intensity.

## V. DISCUSSION

Although the graphene layer is acting only as a resistive reading element of the substrate-intrinsic photopolarization effect, the role of the interface should be discussed. The difference between the FE coercive field (Fig. 1, right inset) and the electric field position of the peak in the electroresistive loop (Fig. 1) indicates the presence of water at the interface. The electrostatic ground state of the interface between the ferroelectric substrate and graphene is therefore verified by warming the sample in vacuum to 350 °C and then cooling it to room temperature. The electroresistive loop is then remeasured at 300 K on the same sample. The both loops initial and after annealing are depicted in Fig. 5.

After annealing, the hysteresis is inverted and electroresistive maxima shrink to the values of the FE coercive force [Fig. 1(b), right inset], as expected for humidity removal from the interface. The electric state obtained after positive poling in darkness (point 1) from 0 to $-7$ kV/cm followed by a sweep from $-7$ to $7$ kV/cm and then back to 0 kV/cm is now lower than that for negative pooling, in agreement with the change of the doping type. The remeasured optical writing of the graphene resistive states is shown in Fig. 6.

The optical-writing and optical-erasing effects still persist after annealing, although they are reversed with respect to Fig. 4(b); light pulses of higher intensity now lead to a decrease in resistance. Therefore, the sample annealing performed here to remove interface humidity confirms the intrinsic nature of the effect. In agreement with the inherent nature of the photo(de)polarization effect of the substrate (see Fig. 2 and Ref. [37]), neither the type of the graphene doping nor the interface humidity undermines the photoresistive performance, although it adds an inverted background. Nevertheless, the control of the interface doping can be potentially used to modify the performance of such photovoltaic-ferroelectric hybrid memory devices. At the same time, it must be underlined that the sample annealing to 350 °C crosses the structural ferroelectric transition of the substrate occurring near 150 °C [45]. Therefore, the annealing affects the structure, and is expected to introduce defects due to large stresses developed near the first-order structural phase transition. Considering the previous discussion, photovoltaic-ferroelectric substrates with higher critical temperatures or a dry 2D layer transfer on the FE are preferable. Additionally, because of the photovoltaic-electrostatic nature of the effect, the remanent polarization must be well defined to ensure a saturated polarization state.

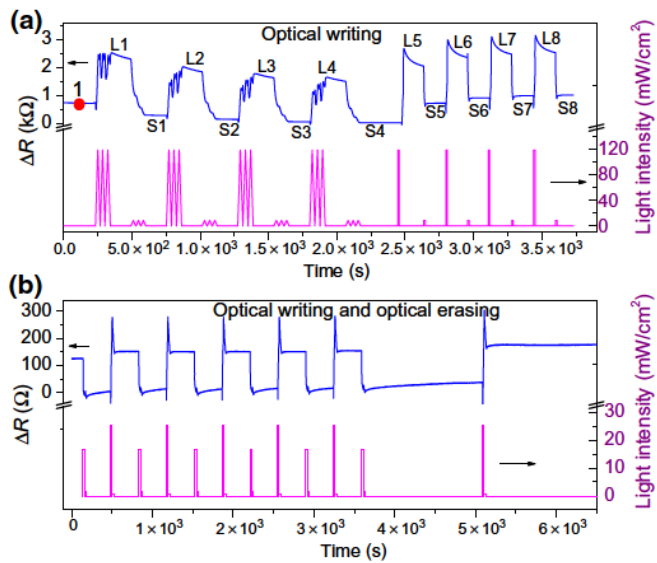

FIG. 4. (a) Time dependence of the graphene resistance illustrating optical writing. The initially positively poled substrate along the [001] direction, state 1 (see the text), is subjected to various light-illumination histories that trigger resistive states of L and S types. (b) States of the S type are written and erased all optically with use of deterministic light pulses. Right scales show corresponding light pulses.

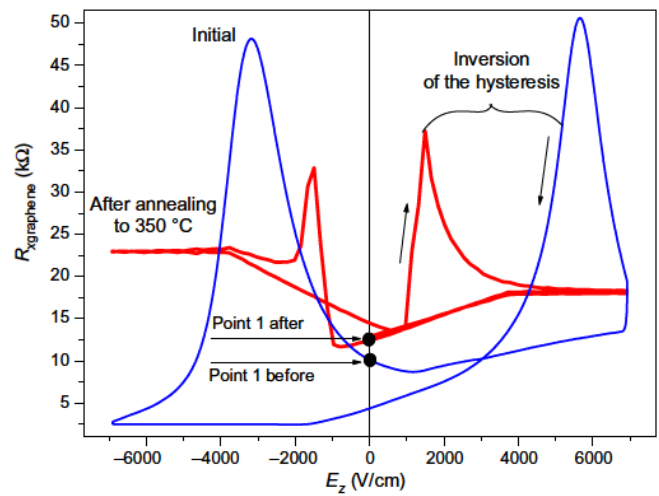

FIG. 5. Room-temperature evolution of the graphene resistance due to the electrostatic doping caused by FE hysteresis for the as-prepared structure and the structure annealed at 350 °C.

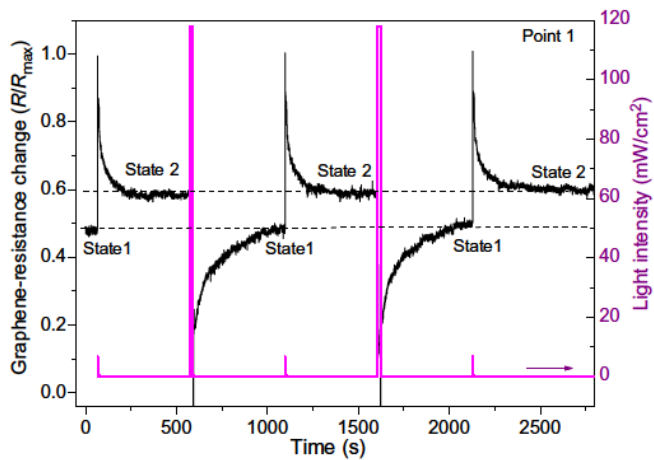

FIG. 6. Optical writing and erasing of graphene resistive states measured at room temperature after annealing of the sample at 350 °C.

## VI. CONCLUSION

By exploiting photovoltaic and FE properties of the substrate, nondestructive optical-writing–optical-erasing cycles are demonstrated. Because of the scarcity of nondestructive optical ways to store information in solids, the results provide a promising concept fundamentally different from the well-known phase-change thermal effect [46–49]. Furthermore, this approach allows all-optical manipulations of the electronic state of 2D overlayers with high responsivity and efficiency, yet importantly conserving the option of nonvolatility. The resistive readout also makes the device operation compatible with resistive random-access memories [50,51], where optical control can be beneficial [52–55]. The reported device also meets miniaturization requirements [56], due to the expected existence of photovoltaic and FE properties down to the nanometer scale. Most importantly, because graphene is highly promising for high-speed optoelectronics [57–59] and plasmonics [60–66], our results also advance its all-optical-control possibility. Such functionality is a key for integrated photonics [67–69] focusing on a number of disruptive quantum-technology applications capable of delivering a source-waveguide-detector platform for quantum-information processing [70–72], sensing, metrology [73], and modulators [74], where compact and scalable electronic-photonic platforms are in great demand.

## ACKNOWLEDGMENTS

This work was partly supported by Labex NIE Grant No. 0058_NIE within the Investissement d'Avenir program ANR-10-IDEX-0002-02. A.S.M. acknowledges the Ph.D. cofund program of the Alsace region.